\documentclass{article}
\usepackage{amsmath,amssymb,amsfonts,euscript} 

\def\qed
{\hfill $\square$}

\def\C{{\mathbb C}}
\def\R{{\mathbb R}}

\def\Z{{\mathbb Z}}

\def\Tend#1#2{\mathop{\longrightarrow}\limits_{#1\rightarrow#2}}

\def\d{{\partial}}

\newtheorem{lem}{Lemma}[section]
\newtheorem{cor}[lem]{Corollary}
\newtheorem{prop}[lem]{Proposition}

\newtheorem{defin}{Definition}

\def\rq
{\noindent {\it Remark.} }

\def\dem
{\noindent {\it Proof.} }

\numberwithin{equation}{section}

\begin{document}
\title{Remarks on nonlinear
Schr\"odinger equations with harmonic potential} 
\author{R\'emi Carles}\date{}
\maketitle
\begin{center}
Math\'ematiques Appliqu\'ees de Bordeaux et UMR 5466 CNRS\\
351 cours de la Lib\'eration\\
33~405~Talence~cedex, France\\
{\tt carles@math.u-bordeaux.fr}
\end{center}
\begin{abstract}
Bose-Einstein condensation is usually modeled by nonlinear
Schr\"odinger equations with harmonic potential. We study the Cauchy
problem for these equations. We show that the local problem can be
treated as in the case with no potential. For the global problem, we
establish an evolution law, which is the analogue of the
pseudo-conformal conservation law for the nonlinear
Schr\"odinger equation. With this evolution law, we give
wave collapse criteria, as well as an upper bound for the blow up
time. Taking the physical scales into account, we  finally give a
lower bound for  the blow up time. 
\end{abstract}
\section{Introduction}
\label{sec:intro}

This paper is devoted to existence and blow up results for the
nonlinear Schr\"odinger equation with isotropic harmonic potential,
\begin{equation}\label{eq:u}
\left\{
\begin{split}
i \hbar \d_t u^\hbar +\frac{\hbar^2}{2} \Delta u^\hbar &
=\frac{\omega^2}{2}x^2 u^\hbar 
+\lambda |u^\hbar|^{2\sigma}u^\hbar ,\ \ \ (t,x)\in \R_+ \times \R^n, \\ 
u^\hbar_{\mid t=0} & = u_0^\hbar,
\end{split}
\right.
\end{equation}
where $\lambda \in \R$, and $\omega, \sigma
>0$.  
Similar equations are considered for Bose-Einstein condensation
(see for instance  \cite{CCT}, \cite{WT99}, \cite{Zhang}), with
$\sigma =1$; the real $\lambda$ may be positive or negative, depending
on the considered chemical element, and is proportional to $\hbar^2$. 
With 
the operators introduced in \cite{CaX01} and \cite{Ca01} (see
Eq.~(\ref{eq:op})), we prove 
existence results which are analogous to the well-known results for
the nonlinear Schr\"odinger equation with no potential 
(see for instance \cite{Caz}). These operators 
simplify the proof of some results of \cite{Oh}, \cite{WT99} and
\cite{Zhang}, as well as the general approach for
(\ref{eq:u}). In  addition, we state two  
evolution laws (Lemma~\ref{lem:e}), which
can be considered as the analogue of the pseudo-conformal evolution
law of the free nonlinear Schr\"odinger field, and allow us to prove
blow up results. Precisely, if we assume that $\lambda$ is negative
(attractive nonlinearity) and $\sigma \geq 2/n$, then under the
condition 
$$\frac{1}{2}\|\hbar \nabla u_0^\hbar \|_{L^2}^2
+\frac{\lambda}{\sigma +1}\|u_0^\hbar \|_{L^{2\sigma +2}}^{2\sigma
+2}<0,$$
the wave collapses at time $t^\hbar_* \leq \frac{\pi}{2\omega}$
(Prop.~\ref{prop:collapse}). Notice that this condition is exactly the
same as the well-known condition for the nonlinear Schr\"odinger
equation with {\it no potential} ($\omega =0$, see e.g. \cite{Caz},
\cite{Sulem}). In particular, blow up occurs for focusing cubic
nonlinearities ($\lambda <0$ and $\sigma =1$)
in space dimensions two and three, but not in space
dimension one. 
Next, we prove that if $\lambda$ is negative and
proportional to $\hbar^2$, $\sigma =1$ (the physical case), and $n=2$
or $3$, then the  wave collapse time can be bounded from {\it
below} by $\frac{\pi}{2\omega} -\Lambda \hbar^\alpha$, for some
constant $\Lambda$ 
and positive number $\alpha$ (Cor.~\ref{cor:below}). 
When $n=1$, we consider the case of a quintic nonlinearity ($\sigma
=2$), which should be the right model for Bose-Einstein Condensation
in low dimension (see \cite{KNSQ}). 
Notice that all
these results are proved for fixed $\hbar$, with constants independent
of $\hbar \in ]0,1]$.

The following quantities
are formally independent of time,
\begin{equation}\label{eq:conservation}
\begin{split}
N^\hbar=& \|u^\hbar(t)\|_{L^2}^2,\\
E^\hbar=& \frac{1}{2}\|\hbar \nabla_x u^\hbar (t)\|_{L^2}^2 
+\frac{\omega^2}{2}\|x u^\hbar(t)\|_{L^2}^2  +
\frac{\lambda}{\sigma +1}
\|u^\hbar(t)\|_{L^{2\sigma+2}}^{2\sigma+2}.
\end{split}
\end{equation} 
If $N^\hbar$ and $E^\hbar$ are
defined at time $t=0$, we prove that the solution $u^\hbar$ is defined locally
in time, with the conservation of $N^\hbar$ and $E^\hbar$, provided
that  $\sigma <2/(n-2)$ when $n\geq 3$.
If $\lambda \geq 0$, then the solution $u^\hbar$ is defined globally in
time. If $\lambda <0$, several cases occur.
\begin{itemize}
\item If $\sigma <2/n$, then the solution is defined globally in
time. 
\item If $\sigma \geq 2/n$, 
then the solution is defined globally in
time if $u_0^\hbar$ is sufficiently small.
\item If $\sigma \geq 2/n$ and $E^\hbar < \frac{\omega^2}{2}\|x
u_0^\hbar\|_{L^2}^2 $, then the solution collapses
at time $t^\hbar_*\leq \frac{\pi}{2\omega}$. 
\end{itemize}

The operators on which our analysis relies are
\begin{equation}\label{eq:op}
J^\hbar_j(t) = \frac{\omega}{\hbar}x_j \sin (\omega t) 
-i \cos (\omega t) \d_j \ ;\ \ \ \  
H^\hbar_j(t)= \omega x_j\cos (\omega t)+i \hbar \sin (\omega t) \d_j. 
\end{equation}
We denote $J^\hbar(t)$ (resp. $H^\hbar(t)$) the operator-valued vector
with components $J^\hbar_j(t)$ (resp. $H_j^\hbar(t)$).
\begin{lem}\label{lem:operators}
$J^\hbar$ and  $H^\hbar$ satisfy the following properties.
\begin{itemize}
\item The commutation relation,
\begin{equation}\label{eq:commut}
\left[J^\hbar(t), i \hbar\d_t +\frac{\hbar^2}{2}\Delta
-\frac{\omega^2}{2} x^2
\right]=\left[H^\hbar(t), i \hbar\d_t +\frac{\hbar^2}{2}\Delta
-\frac{\omega^2}{2} x^2
\right]=0.
\end{equation}
\item Denote $M^\hbar(t)= e^{-i\omega \frac{x^2}{2\hbar}\tan (\omega
t)}$, and  
$Q^\hbar(t) = e^{i\omega \frac{x^2}{2\hbar}\cot (\omega t)}$,
then
\begin{equation}\label{eq:factor}
\begin{split}
J^\hbar (t) &=-i\cos (\omega t) M^\hbar(t)\nabla_x  M^\hbar(-t), \\
H^\hbar (t) &=i\hbar\sin (\omega t) Q^\hbar(t)\nabla_x  Q^\hbar(-t).
\end{split}
\end{equation}
\item The modified Sobolev inequalities. 
For $n\geq 2$, and $2\leq r<  \frac{2n}{n-2}$, define $\delta(r)$ by
\begin{equation}\label{eq:delta}
\delta(r)\equiv n\left(\frac{1}{2}-\frac{1}{r}\right).
\end{equation}
Then for any $2\leq r< \frac{2n}{n-2}$ ($2\leq r \leq \infty$ if
$n=1$), there exists $C_r$ such 
that, 
\begin{equation}\label{eq:sobolev}
\begin{split}
\|v(t)\|_{L^r} & \leq C_r
\|v(t)\|_{L^2}^{1-\delta(r)}
\left( \|J^\hbar(t)v(t)\|_{L^2} +
\|\hbar H^\hbar(t)v(t)\|_{L^2}\right)^{\delta(r)}. 
\end{split}
\end{equation}
\item For any function $F\in C^1(\C, \C)$ of the form $F(z)=zG(|z|^2)$,
we have, 
\begin{equation}\label{eq:jauge}\begin{split}
H^\hbar (t)F(v)&= \d_zF(v)H^\hbar(t)v - \d_{\bar
z}F(v)\overline{H^\hbar(t)v}, \ \forall t \not \in \frac{\pi}{\omega}\Z,\\
J^\hbar(t)F(v)&= \d_zF(v)J^\hbar(t)v - \d_{\bar
z}F(v)\overline{J^\hbar(t)v}, \ \forall t \not \in \frac{\pi}{2\omega}
+ \frac{\pi}{\omega}\Z.
				\end{split}
\end{equation} 
\end{itemize}
\end{lem}
\rq Property (\ref{eq:jauge}) is a  direct
consequence of (\ref{eq:factor}). Property (\ref{eq:sobolev}) is a
consequence of the usual Sobolev inequalities and (\ref{eq:factor}).

\noindent {\bf Notations.} We work with initial data which belong to the space
$$\Sigma :=  \left\{
u\in L^2(\R^n)\ ; \ xu, \nabla u \in L^2(\R^n) \right\}.$$
Notice that $\Sigma = D( \sqrt{-\Delta +|x|^2})$: we work in the same
space as in \cite{Oh}. \\
The notation $r'$
stands for the H\"older conjugate exponent of $r$.

\medskip

\noindent The paper is organized as follows. In Sect.~\ref{sec:exist}, we study
the local Cauchy problem for (\ref{eq:u}), and 
we give sufficient conditions for the solution of (\ref{eq:u}) to be
defined globally in time. In Sect.~\ref{sec:collapse}, we give a
sufficient condition under which the solution blows up in finite
time, and provide an upper bound for the breaking time. In
Sect.~\ref{sec:semi}, we give a lower bound for the breaking time,
that shows that the upper bound underscored 
in Sect.~\ref{sec:collapse} is the physical breaking time  in
the semi-classical limit.

\section{Existence results}
\label{sec:exist}

The solution of  (\ref{eq:u}) with $\lambda =0$ is given by Mehler's
formula (see e.g. \cite{Feyn}),
\begin{equation*}
u^\hbar(t,x)=\left(\frac{\omega}{2i\pi\hbar \sin
\omega t }\right)^{n/2}\int_{\R^n}e^{\frac{i\omega}{ \hbar\sin (\omega t)}
\left(\frac{x^2+y^2}{2}\cos (\omega t) -x.y \right)}u_0^\hbar(y)dy=:
U^\hbar(t)u_0^\hbar(x). 
\end{equation*}
This formula defines a group $U^\hbar(t)$, unitary on $L^2$, for which
Strichartz estimates are available, that is,  mixed
time-space estimates, which are exactly the same as for $U_0^\hbar(t)=
e^{i\frac{t\hbar}{2}\Delta}$. Recall the main properties from which
such estimates stem (see \cite{Caz}, or \cite{KT} for a more general
argument).  
\begin{itemize}
\item The group $U^\hbar(t)$ is  unitary on $L^2$,
$\|U^\hbar(t)\|_{L^2 \rightarrow L^2}=1$.
\item For $ 0<t \leq\frac{\pi}{2\omega}$, the group is dispersive, with
$\|U^\hbar(t)\|_{L^1 \rightarrow L^\infty}\leq C |\hbar t|^{-n/2}$.  
\end{itemize}
We postpone the
precise statement of Strichartz estimates to Sect.~\ref{sec:semi}. 
Duhamel's formula associated to 
(\ref{eq:u}) reads
\begin{equation*}
u^\hbar(t,x) = U^\hbar(t)u_0^\hbar(x) -i\lambda \hbar^{-1}\int_0^t
U^\hbar(t-s)\left( 
|u^\hbar|^{2\sigma}u^\hbar\right)(s,x)ds .
\end{equation*}
Replacing $U^\hbar(t)$ by $U_0^\hbar(t)$ yields Duhamel's formula
associated to   
\begin{equation}\label{eq:u2}
\left\{
\begin{split}
i \hbar\d_t u +\frac{\hbar^2}{2} \Delta u & =\lambda
 |u^\hbar|^{2\sigma}u^\hbar ,\\
u^\hbar_{\mid t=0} & = u_0^\hbar.
\end{split}
\right.
\end{equation}
The local Cauchy problem for this equation is now well-known in many
cases (see for instance \cite{Caz} for a review). In particular, the
local well-posedness in $\Sigma$ is established thanks to the
operators $\hbar\nabla_x$ and $x/\hbar  +it\nabla_x$ (Galilean
operator). This 
result is proved thanks to Strichartz inequalities, and to the
following properties.
\begin{itemize}
\item The above two operators commute with
 $i\hbar\d_t +\frac{\hbar^2}{2}\Delta$.
\item They act on the nonlinearity $|u^\hbar|^{2\sigma}u^\hbar$ like
derivatives.  
\item Gagliardo-Nirenberg inequalities. 
\end{itemize}
From Lemma~\ref{lem:operators}, the operators $H^\hbar$ and $J^\hbar$
meet all 
these requirements. Mimicking the classical proofs for (\ref{eq:u2})
easily yields,
\begin{prop}\label{prop:local}
Let $u_0^\hbar \in\Sigma$. If $n\geq 3$, assume moreover $\sigma <2/(n-2)$. 
Then there exists $T^\hbar >0$ 
such that (\ref{eq:u}) has a unique solution
$u^\hbar \in C([0,T^\hbar],\Sigma)$. Moreover $N^\hbar$ and $E^\hbar$
defined by (\ref{eq:conservation}) are constant for  $t\in [0,T^\hbar]$.
\end{prop}

If $\lambda >0$, the conservations of mass and energy provide {\it a
priori} estimates on the $\Sigma$-norm of $u^\hbar(t)$, and prove global
existence in $\Sigma$. 

If $\lambda <0$ and $\sigma <2/n$, then the
energy $E$ controls the $\Sigma$-norm of $u^\hbar(t)$. Indeed, from
Gagliardo-Nirenberg inequalities (\ref{eq:sobolev}), 
$$\|u^\hbar(t)\|_{L^{2\sigma +2}}\leq C
\|u^\hbar(t)\|_{L^2}^{1-\delta(2\sigma +2)} \left( \|\hbar
J^\hbar(t)u^\hbar\|_{L^2} +
\|H^\hbar (t)u^\hbar\|_{L^2}\right)^{\delta(2\sigma +2)}. $$ 
Notice that the following identity holds point-wise,
\begin{equation*}
|\omega x u^\hbar(t,x)|^2 + |\hbar \nabla_x u^\hbar(t,x)|^2 = 
|\hbar J^\hbar(t)u^\hbar(t,x)|^2 + |H^\hbar(t)u^\hbar(t,x)|^2,
\end{equation*}
and one can rewrite the energy as
\begin{equation}\label{eq:E2}
E^\hbar=\frac{1}{2}\|\hbar J^\hbar(t)u^\hbar\|_{L^2}^2 
+\frac{1}{2}\|H^\hbar(t) u^\hbar\|_{L^2}^2 
+\frac{\lambda}{\sigma +1}\|u^\hbar(t)\|_{L^{2\sigma +2}}^{2\sigma +2}.
\end{equation}
Therefore, using the conservation of mass $N^\hbar$ yields
\begin{equation*}
\|\hbar J^\hbar(t)u^\hbar\|_{L^2}^2 
+\|H^\hbar(t) u^\hbar\|_{L^2}^2 \leq 2 E ^\hbar
+C(\|\hbar J^\hbar(t)u^\hbar\|_{L^2} 
+\|H^\hbar(t) u^\hbar\|_{L^2})^{n\sigma},
\end{equation*}
and if $\sigma <2/n$, then the quantity 
$\|\hbar J^\hbar(t)u^\hbar\|_{L^2}^2 
+\|H^\hbar(t) u^\hbar\|_{L^2}^2$
remains bounded for all times (for any fixed $\hbar$).

Similarly, global existence can be proved for small data.
\begin{prop}\label{prop:global}
Let $u_0^\hbar \in \Sigma$, and if $n\geq 3$, assume
$\sigma <2/(n-2)$.  Then $u^\hbar$ is defined globally in time and
belongs to $C([0,+\infty[,\Sigma)$ in the following cases.
\begin{itemize}
\item $\lambda \geq 0$ (defocusing nonlinearity).
\item $\lambda <0$ (focusing nonlinearity) and $\sigma <2/n$. 
\item $\lambda <0$, $\sigma \geq 2/n$ and $\|u_0^\hbar\|_\Sigma$
sufficiently small. 
\end{itemize}
\end{prop}

\rq In particular, in space dimension one, the solution $u^\hbar$ is
always globally defined for cubic nonlinearities ($\sigma =1$).

\section{Wave collapse}
\label{sec:collapse}
Split the energy $E^\hbar$ into $E_1^\hbar +E_2^\hbar$, with 
\begin{equation*}
\begin{split}
E_1^\hbar(t)& = \frac{1}{2}\|\hbar J^\hbar(t)u^\hbar\|_{L^2}^2 
+\frac{\lambda}{\sigma +1}\cos^2 (\omega t)\|u^\hbar(t)\|_{L^{2\sigma
+2}}^{2\sigma 
+2},\\
E_2^\hbar(t)&=\frac{1}{2}\|H^\hbar(t) u^\hbar\|_{L^2}^2 
+\frac{\lambda}{\sigma +1}\sin^2 (\omega t)\|u^\hbar(t)\|_{L^{2\sigma
+2}}^{2\sigma +2}. 
\end{split}
\end{equation*}
\begin{lem}\label{lem:e}
The quantities $E_1^\hbar$ and $E_2^\hbar$ satisfy the following
evolution laws,
\begin{equation*}
\begin{split}
\frac{d E_1^\hbar}{dt} & = \frac{\omega \lambda}{2\sigma +2}(n\sigma
-2) \sin (2\omega t) \| u^\hbar (t)\|_{L^{2\sigma +2}}^{2\sigma +2},\\
\frac{d E_2^\hbar}{dt} & = \frac{\omega \lambda}{2\sigma +2}
(2-n\sigma)\sin
(2\omega t) \| u^\hbar (t)\|_{L^{2\sigma +2}}^{2\sigma +2}.
\end{split}
\end{equation*}
\end{lem}
\rq This lemma can be regarded as the analogue of the pseudo-conformal
conservation law, discovered by Ginibre and Velo (\cite{GV}) for the
case with no potential ($\omega =0$).

\noindent {\it Sketch of the proof}. 
Expanding $|\hbar J_j^\hbar(t)u^\hbar(t,x)|^2$
yields,
\begin{equation*}
\begin{split}
|\hbar J_j^\hbar(t)u^\hbar(t,x)|^2=& \omega^2 x_j^2 \sin^2(\omega t)
  |u^\hbar(t,x)|^2 + \hbar^2 \cos^2 (\omega t) |\d_j u^\hbar(t,x)|^2
\\
&+ 
\hbar \omega x_j  \operatorname{Im}(\overline{u}\d_j u).
\end{split}
\end{equation*}
When differentiating the above relation with respect to time and
integrating with respect to the space variable, one is
led to computing the following quantities, 
\begin{equation}\label{eq:3eq}
\begin{split}
\d_t \int |x_j u^\hbar(t,x)|^2 dx  =& 2\hbar \operatorname{Im}\int
x_j\overline{u^\hbar}\d_j u^\hbar,\\ 
\d_t \int |\d_j u^\hbar(t,x)|^2 dx =& -2 \frac{\omega^2}{\hbar}
\operatorname{Im}\int
x_j\overline{u^\hbar}\d_j u^\hbar
 -2\frac{\lambda}{\hbar} \operatorname{Im}\int \d_j^2 \overline{u^\hbar}
|u^\hbar|^{2\sigma}u^\hbar,\\
\d_t\operatorname{Im}\int  (x_j\overline{u^\hbar}\d_j u^\hbar)
 =&  \frac{\hbar}{2}\int |\nabla_x u^\hbar |^2
+\frac{\omega^2}{2\hbar} \int x^2 |u^\hbar |^2 +\frac{\lambda}{\hbar}
\int | u^\hbar |^{2\sigma +2}\\
& -\hbar \operatorname{Re}\int x_j \d_j \overline{u^\hbar}\Delta
u^\hbar +\frac{\omega^2}{\hbar}\operatorname{Re}\int x_j \d_j
\overline{u^\hbar}x^2 u^\hbar \\
& + 2\frac{\lambda}{\hbar}
\operatorname{Re}\int x_j \d_j \overline{u^\hbar}
|u^\hbar|^{2\sigma}u^\hbar. 
\end{split}
\end{equation}
It follows,
\begin{equation*}
\begin{split}
\frac{d}{dt}\int |\hbar J^\hbar(t)u^\hbar(t,x)|^2 dx =&
 \frac{\omega\sigma\lambda}{\sigma +1} 
 \sin (2\omega t)\int |u|^{2\sigma +2}\\
&- 2\lambda \hbar \cos^2 (\omega t)
\operatorname{Im}\int \d_j^2 \overline{u} 
|u|^{2\sigma}u.
\end{split}
\end{equation*}
Notice that it is sensible that the right hand side is zero when
$\lambda =0$; from the commutation relation (\ref{eq:commut}), the
$L^2$-norm of $J^\hbar(t)u^\hbar$ is conserved when $\lambda =0$,
since $J^\hbar(t)u^\hbar$ then
solves a linear Schr\"odinger equation.

Finally, the first part of Lemma~\ref{lem:e} follows from the
identity,
$$\frac{d}{dt}\|u^\hbar(t)\|_{L^{2\sigma +2}}^{2\sigma +2}= -\hbar(\sigma
+1)\operatorname{Im}\int |u|^{2\sigma}\overline{u}
\Delta u.$$  
The second part of Lemma~\ref{lem:e} follows from the relation
$E_1^\hbar+E_2^\hbar=E^\hbar=cst$.~\qed

As an application of this lemma, we can prove wave collapse when
$E^\hbar_1(0)<0$. 

\begin{prop}\label{prop:collapse}
Let $u_0^\hbar \in \Sigma$, and if $n\geq 3$, assume
$\sigma <2/(n-2)$. Assume that the nonlinearity is attractive
($\lambda <0$) and $\sigma \geq 2/n$. Assume that
$$\frac{1}{2}\|\hbar \nabla u_0^\hbar \|_{L^2}^2
+\frac{\lambda}{\sigma +1}\|u_0^\hbar \|_{L^{2\sigma +2}}^{2\sigma
+2}<0.$$
Then $u^\hbar$ blows up at time
$t^\hbar_*\leq \pi/2\omega$, 
\begin{equation*}
\exists t^\hbar_*\leq \frac{\pi}{2\omega},\ \ \ 
\lim_{t\rightarrow t^\hbar_*}\|\nabla_x u^\hbar(t)\|_{L^2}=\infty, \ \ \
\textrm{ and }\  
\ \ \lim_{t\rightarrow t^\hbar_*}\| u^\hbar(t)\|_{L^\infty}=\infty. 
\end{equation*}
\end{prop}
\dem From our assumptions, if $u^\hbar \in C([0,T]; \Sigma)$ with $T\leq
\pi/2\omega$, 
\begin{equation}\label{eq:crois}
E_1^\hbar(0) = E^\hbar - \frac{1}{2}\|\omega x 
u_0^\hbar\|_{L^2}^2 < 0
, \textrm{ and }\forall t\in [0,T], \ \ \frac{d E_1^\hbar}{dt} \leq 0
.
\end{equation}
On the other hand, $E_1^\hbar$ can be written as,
\begin{equation*}
\begin{split}
E_1^\hbar(t) =& -\frac{1}{2}\cos (2\omega t) \|\omega x u^\hbar(t,x)\|_{L^2}^2
+ E^\hbar\cos^2 (\omega t)\\
&+\frac{\omega \hbar}{2} \sin (2\omega t)\operatorname{Im} \int
(\overline{u^\hbar}x.\nabla_x u^\hbar). 
\end{split}
\end{equation*}
In particular, Cauchy-Schwarz inequality yields,
\begin{equation*}
\begin{split}
E_1^\hbar(t) \geq & -\frac{1}{2}\cos (2\omega t) \|\omega x
u^\hbar(t,x)\|_{L^2}^2 
+ E^\hbar\cos^2 (\omega t)\\
&-\frac{1}{2}\sin (2\omega
t)\|\omega xu^\hbar(t)\|_{L^2}\|\hbar\nabla_x u^\hbar(t)\|_{L^2}.
\end{split} 
\end{equation*}
So long as $\nabla_x u^\hbar$ remains bounded in $L^2$, so does $x
 u^\hbar$. This follows from the conservations of mass and energy,
 along with Gagliardo-Nirenberg inequality.\\
Assume  $u^\hbar\in C([0,\pi/2\omega]; \Sigma)$. Then letting $t$
go to $\pi/2\omega$ yields 
$$E_1\left(\frac{\pi}{2\omega}\right)\geq \frac{1}{2}
\left\|\omega x 
u^\hbar\left(\frac{\pi}{2\omega},x\right)\right\|_{L^2}^2,$$
which is impossible from (\ref{eq:crois}). 
Thus, there exists $t^\hbar_*\leq \pi/2\omega$ such that 
$$\lim_{t\rightarrow
t^\hbar_*}\|\nabla_x u^\hbar(t)\|_{L^2}=\infty.$$
From the conservation of energy,
$$\lim_{t\rightarrow t^\hbar_*}\|u^\hbar(t)\|_{L^{2\sigma +2}}^{2\sigma +2} =
\infty,$$ 
and the last part of the proposition stems from the conservation of
mass.~\qed  
\medskip

\rq Notice that the blow up condition also reads
$$E^\hbar <\frac{\omega^2}{2}\|x u_0^\hbar \|^2_{L^2}.$$
In term of energy, this means that the blow up occurs for higher values
of the Hamiltonian than in the case with no potential, where the
condition reads $E^\hbar <0$. This sufficient blow up condition varies
continuously with $\omega \geq 0$.

\begin{cor}
Assume $\sigma \geq 2/n$, $\lambda <0$. 
Let $v_0^\hbar \in \Sigma$.  For $k\in \R$, define $u_0^\hbar = k
v_0^\hbar$. Then for 
$|k|$ sufficiently large, $u^\hbar(t,x)$
collapses at time $t^\hbar_* \leq \pi/2\omega$, as in
Prop.~\ref{prop:collapse}. 
\end{cor}
\dem For $|k|$ large, $E^\hbar_1 (0)$ becomes negative, and one can use the
results of Prop.~\ref{prop:collapse}.~\qed

\section{Lower bound for the breaking time}
\label{sec:semi}

In this section, we specify the dependence of the coupling constant
$\lambda$ upon physical constants, and assume 
$\lambda = a \hbar^2.$ We also assume that the nonlinearity is cubic,
$\sigma =1$. 
Physically, $a$ is the $s$-wave scattering length. It is negative in
the case of Bose-Einstein condensation for $\,^7$Li system (\cite{BSTH},
\cite{BSH}). We prove that if the space dimension $n$ is two or three,
then the nonlinear term $a \hbar^2 |u^\hbar|^2
u^\hbar$ in (\ref{eq:u}) is negligible in the semi-classical limit
$\hbar \rightarrow 0$, up to some time depending on $\hbar$. 
This will give us a lower bound for the
breaking time $t^\hbar_*$ when $\hbar \rightarrow 0$, and prove that 
$$t^\hbar_* \Tend \hbar 0 \frac{\pi}{2\omega}.$$
As previously noticed, no blow up occurs for $\sigma =1$ and $n=1$,
that is why we restrict our attention to $n=2$ or $3$.In the
one-dimensional case, it has been proved in \cite{KNSQ} 
that the right model for Bose-Einstein consists in replacing the cubic
nonlinearity $|u^\hbar|^2 u^\hbar$ by the quintic nonlinearity
$|u^\hbar|^4 u^\hbar$. This case is critical for global existence
issues (see Prop.~\ref{prop:global}, Prop.~\ref{prop:collapse}), and
is treated at the end of this section.\\

Define the function $v^\hbar$ as the solution of the linear Cauchy
problem, 
\begin{equation}\label{eq:v}
\left\{
\begin{split}
i \hbar \d_t v^\hbar +\frac{\hbar^2}{2} \Delta v^\hbar &
=\frac{\omega^2}{2}x^2 v^\hbar 
,\\ 
v^\hbar_{\mid t=0} & = u_0^\hbar.
\end{split}
\right.
\end{equation}
\subsection{The case $n=2$ or $3$}
When $n=2$ or $3$, 
recall that we consider now the initial value problem for $u^\hbar$, 
\begin{equation}\label{eq:uh}
\left\{
\begin{split}
i \hbar \d_t u^\hbar +\frac{\hbar^2}{2} \Delta u^\hbar &
=\frac{\omega^2}{2}x^2 u^\hbar 
+a \hbar^2 |u^\hbar|^2 u^\hbar ,\\ 
u^\hbar_{\mid t=0} & = u_0^\hbar,
\end{split}
\right.
\end{equation}
where $a$ is fixed. Our first result is independent of the sign of
$a$. 

\begin{prop}\label{prop:approx}
Assume $n=2$ or $3$. 
Let $u_0^\hbar \in \Sigma$ be such that 
$\|u_0^\hbar\|_{L^2}$, $\|\nabla_x u_0^\hbar\|_{L^2}$ and
$\|x u_0^\hbar\|_{L^2}$
are bounded, \emph{uniformly} with $\hbar \in ]0,1]$. Then there exist
$C,\Lambda,\alpha >0$ and a finite real $\underline q$ such that
\begin{equation*}
\sup_{0\leq t \leq \pi/2\omega-\Lambda \hbar^\alpha} \left\|
A^\hbar (t)( u^\hbar -v^\hbar)(t)\right\|_{L^2} \leq C
\hbar^{1/\underline q} ,
\end{equation*}
where $A^\hbar(t)$ can be either of the operators $Id$, $J^\hbar(t)$
or $H^\hbar(t)$. 
\end{prop}
\rq Notice that the assumption $\|\nabla_x u_0^\hbar\|_{L^2}$ be
bounded uniformly with $\hbar$ means that $u_0^\hbar$ has no
$\hbar$-dependent oscillation . 

From Lemma~\ref{lem:operators}, (\ref{eq:commut}), $J^\hbar v^\hbar$
and $H^\hbar v^\hbar$ solve a linear Schr\"odinger equation with
harmonic potential, and in particular their $L^2$-norms are conserved
with time,
$$\|J^\hbar(t) v^\hbar\|_{L^2}=\|\nabla_x u_0^\hbar\|_{L^2},\ \ 
\|H^\hbar(t) v^\hbar\|_{L^2}=\|\omega x u_0^\hbar\|_{L^2}.$$
We can deduce the following,
\begin{cor}\label{cor:below}
Let $n=2$ or $3$, and $u_0^\hbar \in \Sigma$ be such that 
$\|u_0^\hbar\|_{L^2}$, $\|\nabla_x u_0^\hbar\|_{L^2}$ and
$\|x u_0^\hbar\|_{L^2}$
are bounded, \emph{uniformly} with $\hbar \in ]0,1]$. 
Assume $a<0$ and
$$\| \nabla u_0^\hbar \|_{L^2}^2
+a \|u_0^\hbar \|_{L^4}^4<0.$$
Then there
exists $\Lambda, \alpha>0$ such that
$$\forall \hbar \in ]0,1], \ \ t^\hbar_* \geq \frac{\pi}{2\omega}
- \Lambda \hbar^\alpha.$$
\end{cor}
To prove Prop.~\ref{prop:approx}, we first state precisely the
Strichartz estimates we will use. Recall the
classical definition (see e.g. \cite{Caz}), 
\begin{defin}\label{def:adm}
 A pair $(q,r)$ is \emph{admissible} if $2\leq r
  <\frac{2n}{n-2}$ (resp. $2\leq r\leq \infty$ if $n=1$, $2\leq r<
  \infty$ if $n=2$)  
  and 
$$\frac{2}{q}=\delta(r)\equiv n\left( \frac{1}{2}-\frac{1}{r}\right).$$
\end{defin}
Strichartz estimates 
provide mixed type estimates (that is, in spaces
of the form $L^q_t(L^r_x)$, with $(q,r)$ admissible) of quantities
involving the unitary group 
$$U_0(t)=e^{i\frac{t}{2}\Delta}.$$
A simple
scaling argument yields similar estimates when $U_0$ is replaced with
$e^{i\frac{t\hbar}{2}\Delta}$, with precise dependence upon the
parameter $\hbar$. As noticed in Sect.~\ref{sec:exist}, the same
Strichartz estimates hold when $e^{i\frac{t\hbar}{2}\Delta}$ is
replaced by $U^\hbar(t)$ (provided that only \emph{finite} time
intervals are involved).
\begin{prop}\label{prop:stri}
Let $I$ be a interval contained in $[0,\pi/2\omega]$. 
For any admissible pair $(q,r)$, there exists $C_r$ such that for any
$f\in L^2$, 
\begin{equation*}
    \left\| U^\hbar(t)f\right\|_{L^q(I;L^r)}\leq C_r\hbar^{-1/q}
   \|f\|_{L^2}. 
  \end{equation*}
For any admissible pairs $(q_1,r_1)$ and $(q_2,r_2)$, there exists
$C_{r_1,r_2}$ such that for $F=F(t,x)$, 
\begin{equation}\label{eq:strichnl}
      \left\| \int_{I\cap\{s\leq
      t\}} U^\hbar(t-s)F(s)ds 
      \right\|_{L^{q_1}(I;L^{r_1})}\leq C_{r_1,r_2}\hbar^{-1/q_1 -1/q_2}
      \left\|  F\right\|_{L^{q'_2}(I;L^{r'_2})}. 
    \end{equation}
The above constants are independent of $I\subset [0,\pi/2\omega]$
      and $\hbar \in ]0,1]$. 
\end{prop}
We now state two technical lemmas on which the proof of
Prop.~\ref{prop:approx} relies. 
\begin{lem}\label{lem:alg}
If $n=2$ or $3$, 
there exists $\underline{q}$, $\underline{r}$, $\underline{s}$ and
$\underline{k}$ satisfying
\begin{equation}\label{eq:holder}
  \left\{
  \begin{split}
    \frac{1}{\underline{r}'}&=\frac{1}{\underline{r}}+
\frac{2}{\underline{s}},\\ 
    \frac{1}{\underline{q}'}&=\frac{1}{\underline{q}}+
\frac{2}{\underline{k}},
  \end{split}\right.
\end{equation}
and the additional conditions:
\begin{itemize}
\item The pair $(\underline{q},\underline{r})$ is admissible,
\item $0<\frac{1}{\underline{k}}<\delta(\underline{s})<1$. 
\end{itemize}
\end{lem}
\rq Notice that in particular, $\underline{q}$ is finite. \\
\noindent \emph{Proof of Lemma~\ref{lem:alg}}. 
With
$\delta(\underline{s})=1$, the first part of  (\ref{eq:holder})
becomes 
$$\delta(\underline{r})=\frac{n}{2}-1,$$
and this expression is less than $1$ for 
$n=2$ or $3$. Still with $\delta(\underline{s})=1$, the
second part of (\ref{eq:holder}) yields
$$\frac{2}{\underline{k}}=2-\frac{n}{2},$$
which lies in $]0,2[$ for
$n=2$ or $3$. By continuity, these conditions
are still satisfied for $\delta (\underline{s})$ close to $1$ and $\delta
(\underline{s})<1$. 
\qed

\begin{lem}\label{lem:genest}Assume $n=2$ or $3$, and let $a^\hbar$
solve 
\begin{equation*}
\left\{
\begin{split}
i \hbar \d_t a^\hbar +\frac{\hbar^2}{2} \Delta a^\hbar &
=\frac{\omega^2}{2}x^2 a^\hbar 
+\hbar^2 F^\hbar(a^\hbar)+ \hbar^2 S^\hbar ,\\ 
a^\hbar_{\mid t=0} & = 0.
\end{split}
\right.
\end{equation*}
Assume that there exists $C_0>0$ 
such that for any $t<\pi/2\omega$, 
$$\left\|F^\hbar(a^\hbar)(t)\right\|_{L^{\underline r '}} \leq
\frac{C_0}{\left( \frac{\pi}{2\omega}-t\right)^{2\delta(\underline s)}} 
\left\|a^\hbar(t)\right\|_{L^{\underline r }}.$$
Then there exist $C,\Lambda>0$ independent of $\hbar \in [0,1[$ such that
$$\sup_{0\leq t\leq \frac{\pi}{2\omega}-\Lambda \hbar^\alpha}\|a^\hbar
(t)\|_{L^2} \leq C \hbar^{1-1/\underline q}
 \left\|S^\hbar \right\|_{L^{\underline q '}(0,
\pi/2\omega-\Lambda  \hbar^\alpha; L^{\underline r '})},$$ 
where $\alpha = \frac{1}{\underline k \delta (\underline s)-1}.$
\end{lem}
\noindent \emph{Proof of Lemma~\ref{lem:genest}}. From
(\ref{eq:strichnl}) with $q_1=q_2=\underline q$, for any
$t<\pi/2\omega$, 
\begin{equation}\label{eq:init}
\|a^\hbar\|_{L^{\underline q }(0,t; L^{\underline r })}
\leq 
C \hbar^{1-2/\underline q }\|S^\hbar \|_{L^{\underline q ' }(0,t;
L^{\underline r '})} 
+ C\hbar^{1-2/\underline q }\|F^\hbar(a^\hbar) \|_{L^{\underline q '
}(0,t; L^{\underline r '})}.
\end{equation}
From our assumptions,
$$\|F^\hbar(a^\hbar) \|_{L^{\underline q '
}(0,t; L^{\underline r '})} \leq \left\| \frac{C_0}{\left(
\frac{\pi}{2\omega}-s\right)^{2\delta(\underline s)}}
\|a^\hbar(s)\|_{L^{\underline r}_x}\right\|_{L^{\underline q '
}(0,t)}.$$
Apply H\"older's inequality in time with (\ref{eq:holder}), 
\begin{equation*}
\begin{split}
\|F^\hbar(a^\hbar) \|_{L^{\underline q '
}(0,t; L^{\underline r '})} & \leq C \left( 
\int_0^t \frac{ds}{\left( 
\frac{\pi}{2\omega}-s
\right)^{\underline k \delta ( \underline s)}}
\right)^{2/\underline k}\|a^\hbar\|_{L^{\underline q 
}(0,t; L^{\underline r })}\\
& \leq C\frac{1}{ \left( 
\frac{\pi}{2\omega}-t
\right)^{2\delta ( \underline s)-2/\underline k
}}\|a^\hbar\|_{L^{\underline q  }(0,t; L^{\underline r })}.
\end{split}
\end{equation*}
Plugging this estimate into (\ref{eq:init}) yields, for $t\leq \Lambda
\hbar^\alpha$, 
$$\|a^\hbar\|_{L^{\underline q }(0,t; L^{\underline r })}
\leq 
C \hbar^{1-2/\underline q }\|S^\hbar \|_{L^{\underline q ' }(0,t;
L^{\underline r '})} 
+ C\hbar^{1-2/\underline q }(\Lambda
\hbar^\alpha)^{2/\underline k - 2\delta ( \underline
s)}\|a^\hbar\|_{L^{\underline q  }(0,t; L^{\underline r })}. $$
From (\ref{eq:holder}), the power of $\hbar$ in the last term 
is canceled for
$\alpha = \frac{1}{\underline k \delta (\underline s)-1}$. If in
addition 
$\Lambda$ is sufficiently large, the last term of the above estimate
can be absorbed by the left hand side (up to doubling the constant $C$
for instance), 
\begin{equation*}
\|a^\hbar\|_{L^{\underline q }(0,t; L^{\underline r })}
\leq 
C \hbar^{1-2/\underline q }\|S^\hbar \|_{L^{\underline q ' }(0,t;
L^{\underline r '})} .
\end{equation*}
The last three estimates also imply,
\begin{equation}\label{eq:key}
\|F^\hbar(a^\hbar) \|_{L^{\underline q '
}(0,t; L^{\underline r '})}\leq C \|S^\hbar \|_{L^{\underline q '
}(0,t; L^{\underline r '})} .
\end{equation}
The lemma then follows from Prop.~\ref{prop:stri},
(\ref{eq:strichnl}), with this time $q_1 =\infty$ and
$q_2= \underline q$, along with (\ref{eq:key}).~\qed

\medskip

\noindent \emph{Proof of Proposition~\ref{prop:approx}}. Denote
$w^\hbar = u^\hbar - v^\hbar$ the remainder we want to assess. It
solves the initial value problem,
\begin{equation}\label{eq:w}
\left\{
\begin{split}
i \hbar \d_t w^\hbar +\frac{\hbar^2}{2} \Delta w^\hbar &
=\frac{\omega^2}{2}x^2 w^\hbar + a\hbar^2 |u^\hbar|^2 u^\hbar
,\\ 
w^\hbar_{\mid t=0} & = 0.
\end{split}
\right.
\end{equation}
We first want to apply Lemma~\ref{lem:genest} with $a^\hbar =
w^\hbar$. Since $u^\hbar = v^\hbar +w^\hbar$, we can take
$$F^\hbar(w^\hbar)= a |u^\hbar|^2 w^\hbar, \ \ \ 
S^\hbar = a |u^\hbar|^2 v^\hbar.$$
The point is now to control the $L^{\underline s}$-norm of
$u^\hbar$. Notice that we can easily control the $L^{\underline
s}$-norm of $v^\hbar$. Indeed, as we already emphasized, for any time
$t$, 
$$\|v^\hbar(t)\|_{L^2} = \|u_0^\hbar\|_{L^2},\ \ \ 
\|J^\hbar(t) v^\hbar\|_{L^2} = \|\nabla u_0^\hbar\|_{L^2}.$$
From Lemma~\ref{lem:operators}, (\ref{eq:factor}), and
Gagliardo-Nirenberg inequality, we also have, 
\begin{equation*}
\begin{split}
\|v^\hbar(t)\|_{L^{\underline
s}}& \leq \frac{C}{|\cos (\omega t)|^{\delta( \underline s)} }
 \|v^\hbar(t)\|_{L^2}^{1-\delta( \underline
s)}\|J^\hbar(t) v^\hbar\|_{L^2}^{\delta( \underline s)}  \\
& \leq \frac{C}{\left(\frac{\pi}{2\omega}-t\right)^{\delta( \underline
s)}  }
 \|v^\hbar(t)\|_{L^2}^{1-\delta( \underline
s)}\|J^\hbar(t) v^\hbar\|_{L^2}^{\delta( \underline s)}.
\end{split}
\end{equation*}
Therefore, the assumptions of Prop.~\ref{prop:approx} imply that there 
exists
$C_0>0$ independent of $\hbar$ such that for any $t<\pi/2\omega$, 
\begin{equation*}
\|v^\hbar(t)\|_{L^{\underline
s}} \leq \frac{C_0}{\left(\frac{\pi}{2\omega}-t\right)^{\delta(
\underline s)}  }. 
\end{equation*}
Now $w^\hbar_{\mid t=0}  = 0$ and we know from Prop.~\ref{prop:local}
that there exists $T^\hbar$ such that the $\Sigma$-norm of $w^\hbar$
is continuous on $[0,T^\hbar]$. In particular, there exists $t^\hbar>0$
such that the following inequality,
\begin{equation}\label{eq:tantque}
\|w^\hbar(t)\|_{L^{\underline
s}} \leq \frac{C_0}{\left(\frac{\pi}{2\omega}-t\right)^{\delta(
\underline s)}},
\end{equation}
holds for $t\in [0,t^\hbar]$. So long as (\ref{eq:tantque}) holds, we
have obviously
\begin{equation*}
\|u^\hbar(t)\|_{L^{\underline
s}} \leq \frac{2C_0}{\left(\frac{\pi}{2\omega}-t\right)^{\delta(
\underline s)}  }. 
\end{equation*}
This estimate allows us to apply Lemma~\ref{lem:genest}, which yields,
along with (\ref{eq:holder}), and
provided that $t\leq \pi/2\omega-\Lambda \hbar^\alpha$, 
\begin{equation}\label{eq:1}
\begin{split}
\|w^\hbar\|_{L^\infty(0,t;L^2)}& \leq C \hbar^{1-1/\underline q}\|
|u^\hbar|^2 v^\hbar \|_{L^{\underline q '
}(0,t; L^{\underline r '})}\\
&\leq C \hbar^{1-1/\underline q} \|u^\hbar\|_{L^{\underline k
}(0,t; L^{\underline s})}^2 \|v^\hbar \|_{L^{\underline q 
}(0,t; L^{\underline r })}\\
&\leq C \Lambda^{-2/\underline k}\hbar^{1/\underline q}.
\end{split}
\end{equation}
Now apply the operator $J^\hbar$ to (\ref{eq:w}). From
Lemma~\ref{lem:operators}, $J^\hbar w^\hbar$ solves the same equation
as $w^\hbar$, with $|u^\hbar|^2 u^\hbar$ replaced by
$J^\hbar(|u^\hbar|^2 u^\hbar)$. From (\ref{eq:jauge}), 
$$|J^\hbar(t)(|u^\hbar|^2 u^\hbar)(t,x)|\leq 4 |u^\hbar(t,x)|^2
|J^\hbar(t)u^\hbar(t,x)|.$$
Writing $J^\hbar u^\hbar = J^\hbar v^\hbar + J^\hbar w^\hbar$ and
proceeding as above yields, so long as (\ref{eq:tantque}) holds,
\begin{equation}\label{eq:2}
\|J^\hbar w^\hbar\|_{L^\infty(0,t;L^2)}\leq C \Lambda^{-2/\underline
k}\hbar^{1/\underline q}. 
\end{equation}
Combining (\ref{eq:1}) and (\ref{eq:2}), along with
Gagliardo-Nirenberg inequality, yields, so long as (\ref{eq:tantque})
holds, 
\begin{equation}\label{eq:final}
\|w^\hbar(t)\|_{L^{\underline s}}\leq C\frac{1}{
\left(\frac{\pi}{2\omega} -t
 \right)^{\delta (\underline s)}} \Lambda^{-2/\underline
k}\hbar^{1/\underline q}. 
\end{equation}
Possibly enlarging the value of $\Lambda$, (\ref{eq:final}) shows that
(\ref{eq:tantque}) remains valid up to time $\pi/2\omega-\Lambda
\hbar^\alpha$. This proves Prop.~\ref{prop:approx} when
$A^\hbar(t)=Id$ or $J^\hbar(t)$, from (\ref{eq:1}) and (\ref{eq:2}).
The case $A^\hbar(t)=H^\hbar(t)$ is
then an easy by-product.~\qed 
\medskip

\subsection{The case $n=1$}
We finally prove the analogue of the above results in space dimension
one. When $n=1$, one can do without Strichartz estimates, and simply
use the Sobolev embedding $H^1\subset L^\infty$,
$$\|f\|_{L^\infty}\leq C \|f\|_{L^2}^{1/2}\|\d_x f\|_{L^2}^{1/2}.$$
The wave $u^\hbar$ now solves
\begin{equation}\label{eq:uh1d}
\left\{
\begin{split}
i \hbar \d_t u^\hbar +\frac{\hbar^2}{2} \d^2_x u^\hbar &
=\frac{\omega^2}{2}x^2 u^\hbar 
+a \hbar^2 |u^\hbar|^4 u^\hbar ,\\ 
u^\hbar_{\mid t=0} & = u_0^\hbar.
\end{split}
\right.
\end{equation}
We start with the analogue of Lemma~\ref{lem:genest}.
\begin{lem}
Assume $n=1$, and let $a^\hbar$
solve 
\begin{equation}\label{eq:a}
\left\{
\begin{split}
i \hbar \d_t a^\hbar +\frac{\hbar^2}{2} \d_x^2 a^\hbar &
=\frac{\omega^2}{2}x^2 a^\hbar 
+\hbar^2 F^\hbar(a^\hbar)+ \hbar^2 S^\hbar ,\\ 
a^\hbar_{\mid t=0} & = 0.
\end{split}
\right.
\end{equation}
Assume that there exists $C_0>0$ 
such that for any $t<\pi/2\omega$, 
$$\left\|F^\hbar(a^\hbar)(t)\right\|_{L^2} \leq
\frac{C_0}{\left( \frac{\pi}{2\omega}-t\right)^2} 
\left\|a^\hbar(t)\right\|_{L^2}.$$
Then there exists $C>0$ independent of
$\hbar \in [0,1[$ such that for any $\Lambda\geq 1$, 
$$\sup_{0\leq t\leq \frac{\pi}{2\omega}- \Lambda\hbar}\|a^\hbar
(t)\|_{L^2} \leq C \hbar
 \int_0^{\pi/2\omega- \Lambda\hbar}
\left\|S^\hbar (t)\right\|_{L^2}dt.$$ 
\end{lem}
\dem Multiply (\ref{eq:a}) by  $\overline{a^\hbar}$, integrate with
respect to $x$, and take the imaginary part of the result. This
yields, from Cauchy-Schwarz inequality,
\begin{equation*}
\begin{split}
\frac{d}{dt}\|a^\hbar (t)\|_{L^2} &\leq 2 \hbar
\|F^\hbar(a^\hbar)(t)\|_{L^2} +2 \hbar
\|S^\hbar(t)\|_{L^2} \\
& \leq \frac{2C_0\hbar}{\left( \frac{\pi}{2\omega}-t\right)^2} 
\left\|a^\hbar(t)\right\|_{L^2} + 2 \hbar
\|S^\hbar(t)\|_{L^2}. 
\end{split}
\end{equation*}
The lemma then follows from the Gronwall lemma.~\qed\\

We can now prove the analogue of Prop.~\ref{prop:approx}. 

\begin{prop}\label{prop:approx1d}
Assume $n=1$. 
Let $u_0^\hbar \in \Sigma$ be such that 
$\|u_0^\hbar\|_{L^2}$, $\|\d_x u_0^\hbar\|_{L^2}$ and
$\|x u_0^\hbar\|_{L^2}$
are bounded, \emph{uniformly} with $\hbar \in ]0,1]$. Then there exist
$C,\Lambda >0$ such that
\begin{equation*}
\sup_{0\leq t \leq \pi/2\omega-\Lambda \hbar} \left\|
A^\hbar (t)( u^\hbar -v^\hbar)(t)\right\|_{L^2} \leq C,
\end{equation*}
where $A^\hbar(t)$ can be either of the operators $Id$, $J^\hbar(t)$
or $H^\hbar(t)$. 
\end{prop}
\dem The proof follows the proof of Prop.~\ref{prop:approx} very
closely, if we take $\underline q =\infty$, $(\underline s,\underline
k)= ( \infty, 4)$. 
Denote
$w^\hbar = u^\hbar - v^\hbar$ the remainder we want to assess. It
solves the initial value problem,
\begin{equation*}
\left\{
\begin{split}
i \hbar \d_t w^\hbar +\frac{\hbar^2}{2} \d_x^2 w^\hbar &
=\frac{\omega^2}{2}x^2 w^\hbar + a\hbar^2 |u^\hbar|^4 u^\hbar
,\\ 
w^\hbar_{\mid t=0} & = 0.
\end{split}
\right.
\end{equation*}
We first want to apply the above lemma with $a^\hbar =
w^\hbar$. Since $u^\hbar = v^\hbar +w^\hbar$, we can take
$$F^\hbar(w^\hbar)= a |u^\hbar|^4 w^\hbar, \ \ \ 
S^\hbar = a |u^\hbar|^4 v^\hbar.$$
The point is now to control the $L^\infty$-norm of
$u^\hbar$. Notice that we can easily control the $L^\infty$-norm of
$v^\hbar$. Indeed, as we already emphasized, for any time 
$t$, 
$$\|v^\hbar(t)\|_{L^2} = \|u_0^\hbar\|_{L^2},\ \ \ 
\|J^\hbar(t) v^\hbar\|_{L^2} = \|\d_x u_0^\hbar\|_{L^2}.$$
From Lemma~\ref{lem:operators}, (\ref{eq:factor}), and
Gagliardo-Nirenberg inequality, we also have, 
\begin{equation*}
\begin{split}
\|v^\hbar(t)\|_{L^\infty}& \leq \frac{C}{|\cos (\omega t)|^{1/2} }
 \|v^\hbar(t)\|_{L^2}^{1/2}
\|J^\hbar(t) v^\hbar\|_{L^2}^{1/2}  \\
& \leq \frac{C}{\left(\frac{\pi}{2\omega}-t\right)^{1/2}}
 \|v^\hbar(t)\|_{L^2}^{1/2}
\|J^\hbar(t) v^\hbar\|_{L^2}^{1/2}.
\end{split}
\end{equation*}
Therefore, the assumptions of Prop.~\ref{prop:approx1d} imply that there 
exists
$C_0>0$ independent of $\hbar$ such that for any $t<\pi/2\omega$, 
\begin{equation*}
\|v^\hbar(t)\|_{L^\infty} 
\leq \frac{C_0}{\left(\frac{\pi}{2\omega}-t\right)^{1/2}  }. 
\end{equation*}
So long as
\begin{equation}\label{eq:tantque1d}
\|w^\hbar(t)\|_{L^\infty} \leq
\frac{C_0}{\left(\frac{\pi}{2\omega}-t\right)^{1/2}},
\end{equation}
holds, we
have obviously
\begin{equation*}
\|u^\hbar(t)\|_{L^\infty} \leq
\frac{2C_0}{\left(\frac{\pi}{2\omega}-t\right)^{1/2}  }. 
\end{equation*}
This estimate allows us to apply the above lemma, which yields,
provided that $t\leq \pi/2\omega-\Lambda \hbar$, 
\begin{equation}\label{eq:11d}
\begin{split}
\|w^\hbar\|_{L^\infty(0,t;L^2)}& \leq C \hbar\|
|u^\hbar|^4 v^\hbar \|_{L^\infty (0,t; L^2)}\\
&\leq C \hbar \|u^\hbar\|_{L^4(0,t; L^\infty)}^2 \|v^\hbar
\|_{L^\infty(0,t; L^2)}\\  
&\leq C \Lambda^{-1}.
\end{split}
\end{equation}
Similarly, applying the operator $J^\hbar$ to (\ref{eq:w}) yields, so long as
(\ref{eq:tantque}) holds, 
\begin{equation}\label{eq:21d}
\|J^\hbar w^\hbar\|_{L^\infty(0,t;L^2)}\leq C \Lambda^{-1}.  
\end{equation}
Combining (\ref{eq:11d}) and (\ref{eq:21d}), along with
Gagliardo-Nirenberg inequality, yields, so long as (\ref{eq:tantque1d})
holds, 
\begin{equation}\label{eq:final1d}
\|w^\hbar(t)\|_{L^\infty}\leq C\frac{1}{
\left(\frac{\pi}{2\omega} -t
 \right)^{1/2}} \Lambda^{-1}. 
\end{equation}
Taking $\Lambda$ large enough, (\ref{eq:final1d}) shows that
(\ref{eq:tantque1d}) remains valid up to time $\pi/2\omega-\Lambda
\hbar$. This proves Prop.~\ref{prop:approx1d} when
$A^\hbar(t)=Id$ or $J^\hbar(t)$, from (\ref{eq:11d}) and (\ref{eq:21d}).
The case $A^\hbar(t)=H^\hbar(t)$ is
then an easy by-product.~\qed 

\begin{cor}\label{cor:below1d}
Let $n=1$, and $u_0^\hbar \in \Sigma$ be such that 
$\|u_0^\hbar\|_{L^2}$, $\|\d_x u_0^\hbar\|_{L^2}$ and
$\|x u_0^\hbar\|_{L^2}$
are bounded, \emph{uniformly} with $\hbar \in ]0,1]$. 
Assume $a<0$ and
$$\frac{1}{2}\| \d_x u_0^\hbar \|_{L^2}^2
+\frac{a}{3}\|u_0^\hbar \|_{L^6}^6<0.$$
Then there
exists $\Lambda>0$ such that
$$\forall \hbar \in ]0,1], \ \ t^\hbar_* \geq \frac{\pi}{2\omega}
- \Lambda \hbar.$$
\end{cor}
\noindent\emph{Acknowledgment}. The results in this paper were
improved thanks to remarks made by T. Colin.

\bibliographystyle{amsplain}
\bibliography{carles}
\end{document}